# Stark effect in low-dimension hydrogen

Thomas Garm Pedersen

*Department of Physics and Nanotechnology, Aalborg University, DK-9220 Aalborg Øst, Denmark*

Héctor Mera and Branislav K. Nikolić

*Department of Physics and Astronomy, University of Delaware, Newark, DE 19716-2570, USA*

Studies of atomic systems in electric fields are challenging because of the diverging perturbation series. However, physically meaningful Stark shifts and ionization rates can be found by analytical continuation of the series using appropriate branch cut functions. We apply this approach to low-dimensional hydrogen atoms in order to study the effects of reduced dimensionality. We find that modifications by the electric field are strongly suppressed in reduced dimensions. This finding is explained from a Landau-type analysis of the ionization process.

## 1. Introduction

Atomic systems placed in electrostatic fields have played a central role in applications of quantum mechanics and semi-classical physics. For the hydrogen atom, early work demonstrated that a finite order perturbation analysis provides well-defined (hyper-)polarizabilities of a given initial state [1,2]. However, the perturbation series reached by expansion in powers of the field strength has, in fact, zero radius of convergence [3-5]. Hence, the unboundedness of the perturbation makes a non-perturbative mathematical analysis challenging. Physically, a strong electric field manifests itself in the form of ionization and energy (Stark) shifts. This may be viewed mathematically as replacing real-valued energies by complex resonances, in which the imaginary part determines the ionization rate [5]. Such resonances can be found non-perturbatively by matching the wave function to the proper asymptotic solution far from the atom [6-8]. In spite of the diverging series, physically meaningful ionization rates and energy shifts can be obtained from the perturbation expansions using Borel resummation [4,5,9-11]. However, Borel resummation for this problem requires a large number of terms (~70) of the divergent perturbative series as an input. Recently [12], we have proposed a very efficient alternative based on matching a class of analytical continuations functions to the first few terms in the perturbation expansion. In practice, this class was taken to be Gauss hypergeometric functions that have branch cuts and, thereby, may produce a complex result even if a real-value field is supplied as input. The imaginary part is then the ionization rate. Choosing the $_2F_1$ class of hypergeometric functions, only the four lowest terms in the expansion are required. Nevertheless, excellent agreement with highly accurate, but much more demanding, approaches was demonstrated.

In the present work, we aim at applying this approach to evaluate Stark shifts and ionization rates of a broader class of quantum systems, viz. hydrogen-like systems in arbitrary dimensional space. One- and two-dimensional hydrogen atoms have been studied in different contexts [13,14] and used to highlight the effects of reduced dimensionality. In fact, low-dimensional hydrogenic systems are realized in nature in the form of electron-hole pairs ("excitons") in low-dimensional quantum structures such as quantum wells, quantum wires, and carbon nanotubes [15-18]. For such nanostructures, non-integer dimensions $\alpha$ are often considered [15-17] in order to describe quantum wells of finite thickness ($2 < \alpha < 3$) and nanowires or –tubes of finite cross section ($1 < \alpha < 2$). Thus, $\alpha$-dimensional hydrogenic models have experimental relevance as well. The question we ask in the following is: What is the effect of reduced dimensionality on Stark shifts and ionization rates? We clearly expect tighter confinement to counteract the electric field but precisely by how much is not known. By formulating our model system as a hydrogenic system with an arbitrary reduced dimensionality, we can quite generally compute the dependence of e.g. ionization rates on dimension. Hence, the above question can be given a precise, quantitative answer. In the process of the analysis, we will enlarge the class of analytical continuation functions. The low-order perturbation expansion required to fix these functions is obtained using an extension of the work of Privman [3]. Hence, we formulate the $\alpha$- dimensional eigenvalue problem in the presence of an electrostatic field in terms of parabolic coordinates [3] and solve order by order through iteration. We restrict the analysis to the ground state but extension to excited states is certainly possible.

## 2. Model and Perturbation Analysis

A hydrogenic atom placed in an electrostatic field $\vec{\mathcal{E}} = \mathcal{E}\hat{z}$ in an $\alpha$- dimensional space is described by the eigenvalue problem

$$\left\{ -\frac{1}{2}\nabla_\alpha^2 - \frac{1}{r} + \mathcal{E}z \right\}\psi = E\psi. \qquad (1)$$

Here, $\nabla_\alpha^2$ is the $\alpha$- dimensional Laplacian and natural units are adopted throughout the paper using the reduced mass $\mu$ of the two-particle system as the unit of mass: $e = 4\pi\varepsilon_r\varepsilon_0 = \hbar = \mu = 1$ with $\varepsilon_r$ the relative dielectric constant of the ambient medium. We are assuming translational invariance along at least one extended dimension and, hence, only $\alpha \geq 1$ makes sense. In the presence of the field, the natural coordinates are the parabolic ones. We restrict the analysis to states that are rotationally symmetric around the field axis $\hat{z}$. The Laplacian for arbitrary integer-dimensional space was derived in Ref. [19]. However, starting from the usual $\alpha$- dimensional Laplacian in spherical coordinates [15-17], it is readily demonstrated that the expression is valid in non-integer dimensions as well. Hence, we introduce $\xi = r + z$ and $\eta = r - z$ and with $p = \frac{\alpha-1}{2}$ find

$$\nabla_\alpha^2 = \frac{4}{\xi+\eta}\left\{\frac{1}{\xi^{p-1}}\frac{\partial}{\partial\xi}\xi^p\frac{\partial}{\partial\xi} + \frac{1}{\eta^{p-1}}\frac{\partial}{\partial\eta}\eta^p\frac{\partial}{\partial\eta}\right\}. \qquad (2)$$

Similarly, the potential energy terms are $-1/r = -2/(\xi + \eta)$ for the Coulomb potential and $\mathcal{E}z = \frac{1}{2}\mathcal{E}(\xi - \eta)$ for the electrostatic potential. This allows us to reformulate the eigenvalue problem as

$$\left\{\frac{1}{\xi^{p-1}}\frac{\partial}{\partial\xi}\xi^p\frac{\partial}{\partial\xi} + \frac{1}{\eta^{p-1}}\frac{\partial}{\partial\eta}\eta^p\frac{\partial}{\partial\eta} - \mathcal{E}\frac{\xi^2-\eta^2}{4} + E\frac{\xi+\eta}{2} + 1\right\}\psi = 0. \quad (3)$$

We now follow Privman [3] in that we apply logarithmic perturbation theory [20] and introduce (i) $\beta = (E/E_0)^{1/2}$ with $E_0 = -\frac{1}{2p^2}$ the unperturbed ground state energy, (ii) a scaled field strength $F = \mathcal{E}/4\beta^3$ and parabolic coordinates $x = \beta\xi$ and $y = \beta\eta$, and (iii) a set of separation constants $\beta_1$ and $\beta_2$ satisfying $\beta_1 + \beta_2 = 1/\beta$. Thus, writing $\psi = f(x)g(y)$ we find two decoupled eigenvalue problems

$$\left\{\frac{1}{x^{p-1}}\frac{\partial}{\partial x}x^p\frac{\partial}{\partial x} + \beta_1 - Fx^2 - \frac{x}{4p^2}\right\}f = 0,$$

$$\left\{\frac{1}{y^{p-1}}\frac{\partial}{\partial y}y^p\frac{\partial}{\partial y} + \beta_2 + Fy^2 - \frac{y}{4p^2}\right\}g = 0. \quad (4)$$

As in the 3D case, these only differ mathematically by the sign of the field term. Focusing on the first of these, we then define $z(x) \equiv d\ln f / dx$ and expand in Taylor series

$$\beta_1 = \sum_{n=0}^{\infty} a_n F^n, \quad z(x) = \sum_{n=0}^{\infty} z_n(x) F^n. \quad (5)$$

The rest of the calculation proceeds by solving order by order, keeping $z$ regular at the origin. In $\alpha$-dimensional space, the unperturbed ground state is $f_0(x) = N\exp\left(-\frac{x}{\alpha-1}\right)$ with energy $E_0 = -\frac{1}{2p^2}$. Hence, $z_0(x) = 1/(1-\alpha)$ and by symmetry $a_0 = 1/2$. To illustrate the general approach, we note that upon collecting first order terms, $z_1(x)$ obeys the condition

$$\left\{\frac{p}{x} + 2z_0(x)\right\}z_1(x) + \frac{dz_1(x)}{dx} = -\frac{a_1}{x} + x. \quad (6)$$

Requiring regularity at infinity leads to a solution of the form

$$z_1(x) = \frac{x^{-p}}{f_0^2(x)}\int_x^{\infty} t^{p-1}f_0^2(t)(a_1 - t^2)dt. \quad (7)$$

Now, if we require regularity at the *origin* as well, it follows that the unknown $a_1$ must be determined by

$$a_1 = \frac{\int_0^\infty x^{p+1} f_0^2(x)dx}{\int_0^\infty x^{p-1} f_0^2(x)dx} = \frac{(\alpha+1)(\alpha-1)^3}{16}. \tag{8}$$

Continuing to successively higher orders, we find for $k > 1$

$$a_k = -\frac{\int_0^\infty x^p f_0^2(x) \sum_{i=1}^{k-1} z_i(x) z_{k-i}(x) dx}{\int_0^\infty x^{p-1} f_0^2(x) dx}. \tag{9}$$

A similar approach can be followed to compute $\beta_2 = \sum_{n=0}^\infty b_n F^n$. However, as the equations Eq.(4) for $f$ and $g$ only differ by the sign of $F$ it follows immediately that $b_n = (-1)^n a_n$. Eventually, the condition for the separation constants then becomes $1/\beta = 2\sum_{n=0}^\infty a_{2n} F^{2n}$. Computing the series to sufficiently high order and solving for the energy $E = \beta^2 E_0$ produces the desired perturbation series for the energy in powers of the electrostatic field $E(\mathcal{E}) = \sum_{n=0}^\infty E_{2n} \mathcal{E}^{2n}$. The coefficients follow the form $E_{2n} = -(\alpha+1)((\alpha-1)/4)^{6n-2} F_{2n}(\alpha)$ with $F_{2n}(\alpha)$ a polynomial of degree $2n-1$. For $n=1$ to 4, these are given in Tab. 1. It is readily verified that the general result agrees with the known cases $\alpha = 3$ [3] and $\alpha = 2$ [14]. Moreover, for arbitrary $\alpha$ the result for $E_2 = -(\alpha+1)((\alpha-1)/4)^4 (2\alpha+3)$ agrees with the polarizability found in Ref. [17]. Note that all terms vanish if $\alpha = 1$ as a consequence of the pathological nature of the strictly one-dimensional Coulomb problem [13,18], for which the delta-function localized ground state is not polarizable.

| $n$ | $F_{2n}(\alpha)$ |
|---|---|
| 1 | $2\alpha + 3$ |
| 2 | $96\alpha^3 + 645\alpha^2 + 1522\alpha + 1257$ |
| 3 | $2(5888\alpha^5 + 79573\alpha^4 + 453872\alpha^3 + 1361778\alpha^2 + 2139416\alpha + 1399473)$ |
| 4 | $2031616\alpha^7 + 43604973\alpha^6 + 423670118\alpha^5 + 2410476263\alpha^4 + 8642479892\alpha^3 + 19432592955\alpha^2 + 25222378022\alpha + 14478766161$ |

Table 1. First four polynomials in the perturbation series for the $\alpha$-dimensional Stark problem.

## 3. Hypergeometric Resummation

We wish to exploit our recently developed resummation technique to extract physical properties like resonances at arbitrary field strength from low-order perturbations series generated assuming weak electric fields. As detailed in Ref. [12], the fundamental idea is that the low-order series is regarded as the first few terms in a Taylor series of an analytic continuation function with a suitable branch cut. This property ensures that the imaginary part of the resonance, i.e. the ionization rate, is obtained following the continuation procedure. Gauss hypergeometric functions $_2F_1$ were selected for this purpose and shown to lead to good agreement with existing approaches for, e.g., the 3D hydrogen Stark problem. We therefore aim to apply the hypergeometric resummation technique to the low-dimensional case in the present work.

Before turning to this application, we wish to address a particular issue related to the branch cut structure of $_2F_1$, however. When expanded around $z=0$, the function is defined by

$$_2F_1(h_1,h_2,h_3,z) = 1 + \frac{h_1 h_2 z}{h_3} + \frac{h_1(1+h_1)h_2(1+h_2)z^2}{2h_3(1+h_3)} + \frac{h_1(1+h_1)(2+h_1)h_2(1+h_2)(2+h_2)z^3}{6h_3(1+h_3)(2+h_3)} + \ldots \quad (10)$$

The approach in Ref. [12] was based on writing $z = h_4(\mathcal{E}/4)^2$ and determining the four coefficients $h_{1-4}$ by matching to the fourth order perturbation series. While this leads to a well-behaved and accurate result for intermediate field strengths it is bound to fail for small fields, however. The reason is that the branch cut runs between $z=1$ and $z=\infty$ and, therefore, necessarily produces a real-valued result when the function is evaluated at an argument $0 \leq z < 1$. While this may be acceptable physically, because the actual imaginary part is exceedingly small for $z<1$, it is nevertheless not entirely satisfactory. Thus, it is tempting to consider instead the slightly modified class of functions $_2F_1(h_1,h_2,h_3,1+z)$ defined with a shifted argument. This class would ensure a finite imaginary part at all field strengths. Unfortunately, this class of functions does not have a simple Taylor series when expanded around $z=0$. Rather, powers of the form $z^{h_3-h_1-h_2+m}$, with $m$ a non-negative integer, appear in addition to a regular series. To ensure the correct low-$z$ behavior, we consequently require $h_3 = h_1 + h_2 + l$, where $l$ is a fixed integer. In this manner, one finds

$$_2F_1(h_1,h_2,h_1+h_2+l,1+z) = \frac{\Gamma(l+h_1+h_2)}{\Gamma(l+h_1)\Gamma(l+h_2)}\{F_0(z) + F_l(z)\}, \quad (11)$$

where

$$F_0(z) = \Gamma(l) + h_1 h_2 \Gamma(l-1)z + \frac{1}{2}h_1(1+h_1)h_2(1+h_2)\Gamma(l-2)z^2 + \ldots \quad (12)$$

and the lowest power found in the Taylor expansion of $F_l(z)$ is $z^l$. It follows that only the normal series $F_0(z)$ needs to be considered when matching to the fourth order perturbation series provided a power $l > 4$ is used. In fact, provided this condition is obeyed, $l$ does not have to be integer. In practice, we find that the form

$$E(\mathcal{E}) = E_0 \left\{ 1 + h_4 z \frac{\Gamma(l+h_1)\Gamma(l+h_2)}{\Gamma(l+h_1+h_2)} {}_2F_1(h_1, h_2, h_1+h_2+l, 1+h_3 z) \right\}, \quad (13)$$

with $z = (\mathcal{E}/4)^2$, is highly suited for the present purpose. Note that upon separating out the $E_0$ term and factoring out $h_4 z$ in the remainder, we obtain the desired form having precisely four unknowns $h_{1-4}$ as before. This is the form we match to the perturbation series below. We separate the complex resonance Eq.(13) into real and imaginary parts. These are, respectively, the Stark energy $\Delta$ and half the ionization decay rate $\Gamma$, i.e. $E(\mathcal{E}) = \Delta - i\Gamma/2$. The results are relatively insensitive to the value of $l$ as long as $l \gg 4$. As a practical strategy, we use known exact data for the 3D case [8] to select the best value and, in this manner, a value of $l = 30$ has been selected for the numerical routine. A comparison between the hypergeometric result and exact data is shown in Fig. 1. Throughout the entire range of field strengths $0 \leq \mathcal{E} \leq 1$, a remarkable agreement is observed. Note, that the minimum in Stark energy around $\mathcal{E} \approx 0.7$ is reproduced. Hence, in addition to the non-vanishing decay rate at all field strengths ensured by our modified hypergeometric ansatz, we also improve overall agreement for large fields as compared to our original approach [12].

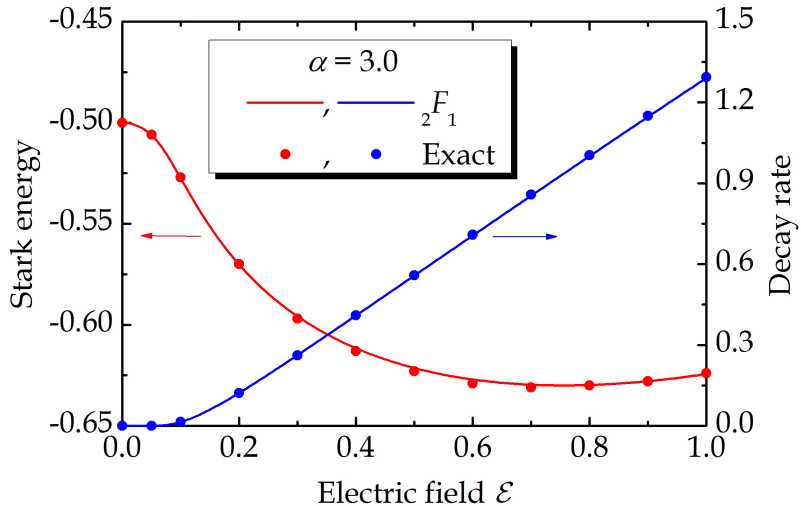

Figure 1. Stark energy (left axis) and decay rate (right axis) of three dimensional hydrogen in the present approach and in comparison to exact values from Ref. [8].

In Fig. 2, we plot results for the Stark energy and decay rate for a range of integer and fractional dimensions $\alpha = 3, 2.5, 2$, and $1.5$ corresponding to $p = 1, 0.75, 0.5$, and $0.25$. As expected, the zero-field limit coincides with the unperturbed result $E_0 = -\frac{1}{2p^2}$. At increased field strengths, the Stark energy decreases and develops a "knee" structure, beyond which the slope decreases. The decay rate is highly suppressed at low fields but increases nearly linearly with field strength above a certain critical point. Fitting the slope, it is found to vary approximately as $p^\gamma$ with $\gamma \approx 1.4$. Obviously, the decay rate decreases rapidly as the dimensionality is reduced. The intersection of the linear approximation with the field axis provides a measure of the critical turn-on field strength. For the four cases studied, the critical fields are 0.12, 0.33, 1.3, and 10.2, respectively. Thus, upon reducing the dimension from 3 to 1.5, the field required to effectively ionize the atom increases by nearly two orders of magnitude. The validity of the computed decay rates can be ascertained using the connection between field dependent decay rate $\Gamma(\mathcal{E})$ and the original perturbation coefficients $E_{2n}$ [10]

$$E_{2n} = -\frac{1}{\pi}\int_0^\infty \frac{\Gamma(\mathcal{E})}{\mathcal{E}^{2n+1}} d\mathcal{E}, \tag{14}$$

valid for $n > 1$. For all dimensionalities studied here, we find that this condition is obeyed to a very high degree of accuracy. This testifies further to the soundness of the approach.

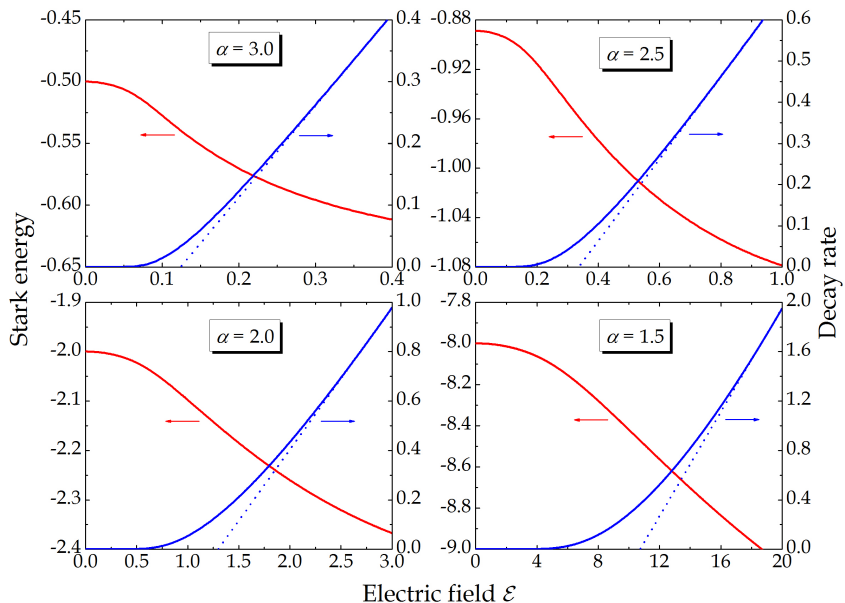

Figure 2. Energy and decay rate as a function of field strength for a range of integer and non-integer dimensions. Note the different field scales. The dashed lines are linear fits to the high-field decay rates.

A simple analytical estimate of the decay rate can be obtained using a modified Landau approach [21]. To this end, we inspect the y-equation in Eq.(4). Writing $g(y) = y^{-p/2}\chi(y)$ we find $-\chi''(y) + U\chi(y) = 0$ with

$$U(y) \approx -\frac{1}{4}\left(\frac{2}{y} + \mathcal{E}y - \frac{1}{p^2} + \frac{p(2-p)}{y^2}\right), \tag{15}$$

where the low-field limit $\beta_2 \approx 1/2$ and $\beta \approx 1$ is assumed. We now compute the WKB transmittance $T = \exp\{-2\int_{y_1}^{y_2} \sqrt{-U}dy\}$ between the classical turning points $y_{1,2}$. Adapting the Landau approach [21] to the $\alpha$-dimensional case then yields

$$T \approx \left(\frac{4}{p^2 \mathcal{E} y_1}\right)^p \exp\left\{-\frac{2}{3p^3 \mathcal{E}}\right\} e^{y_1/p}. \qquad (16)$$

This result agrees with the usual 3D case as is easily seen by taking $p = 1$. Moreover, the $p^3$ factor in the exponential is readily explained by the general field factor [22] $\exp\{-2(2I_P)^{3/2}/(3\mathcal{E})\}$, with a modified ionization potential $I_P = \frac{1}{2p^2}$. The result means, however, that we generally expect the low-field ionization rate to vary with dimension as $\Gamma \propto \exp\{-2/(3p^3 \mathcal{E})\}/\mathcal{E}^p$. In Fig. 3, this is indeed seen to be the case. There, the decay rate is plotted on a logarithmic scale and compared to the Landau-type expression. In particular, good agreement is found for low field strengths. For larger fields, the Landau fit tends to overestimate the decay rate. The discrepancy, however, is quite small for the reduced dimensions $\alpha = 2$ and $1.5$. The agreement with the Landau expression allows us to quantify the suppressed decay rate with reduced dimension. Thus, the dominant factor is the exponential, from which it appears that the effective field in low-dimensional geometries is reduced from $\mathcal{E}$ to $p^3 \mathcal{E}$. Consequently, going from $\alpha = 3$ to $\alpha = 1.5$ effectively reduces the field strength by a factor $4^3 = 64$ in good agreement with the nearly two orders of magnitude increase in the critical field found above.

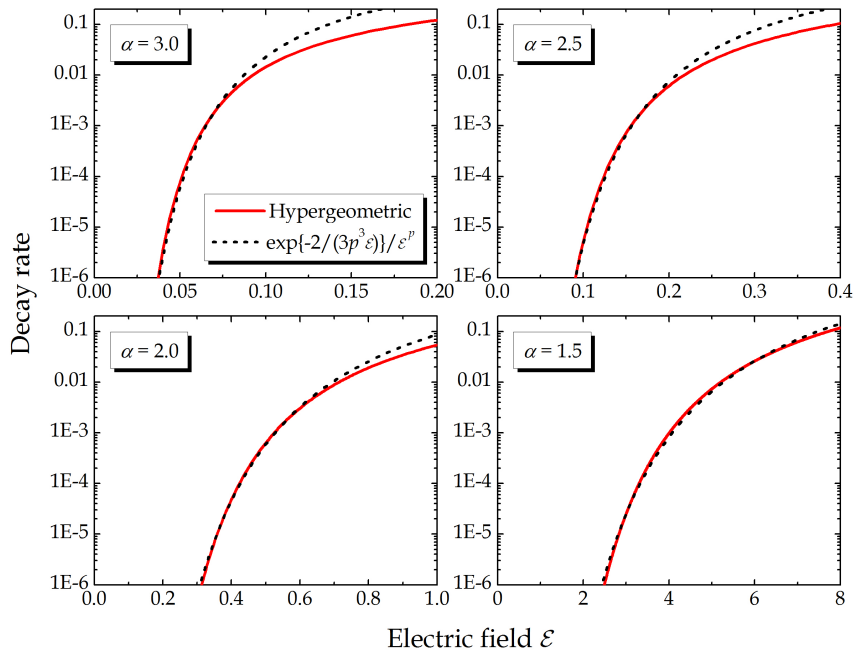

Figure 3. Field dependence of the decay rate for various dimensions. The solid red lines are the hypergeometric resummation results and the dashed black lines are Landau-type fits.

## 4. Summary

In summary, we have applied a recently proposed hypergeometric resummation technique to the study of low-dimensional hydrogen atoms in strong electrostatic fields. In this way, the effect of reduced dimensionality on Stark shifts and ionization decay rates has been identified. We have introduced an enlarged class of analytical continuation functions that ensure a non-vanishing decay rate at arbitrarily small field strengths. For the three-dimensional case, excellent agreement with exact results is demonstrated. Upon reducing the dimension from 3 to 1.5, the critical field required for strong ionization is increased by nearly two orders of magnitude. This finding is explained by a Landau-type analysis adapted to the low-dimensional geometry.

## Acknowledgement

We thank Lars B. Madsen for useful comments. TGP is supported by the QUSCOPE center funded by the Villum foundation. HM and BKN are supported by NSF under Grant No. ECCS 1509094.